# Architected Structural Material Design Inspired by Diatoms: Merging Nature's Beauty With Engineering Through Biomimetics

L. Musenich, F. Libonati*

University of Genoa, Polytechnic School, Department of Mechanical, Energy, Management and Transportation Engineering, Via all'Opera Pia 15/A, 16145, Genova, Italy

*Corresponding author: flavia.libonati@unige.it

## Abstract

Imagine a world where beauty and technology move in perfect harmony, revealing tiny masterpieces hidden in the water. This world is Nature, and diatoms are the stars. Masters of assembling complex hierarchical glass structures, these microscopic algae are true technological gems, and their potential is boundless when it comes to transforming our world. From pharmaceutical applications to biomaterial synthesis, water purification to advanced optical systems, solar cells to the production of new fuels, architecture to design objects—diatoms teach us that Nature's potential is limitless, proving to be invaluable allies for innovation and sustainable development. But how can organisms so small hold such immense potential? In this chapter, we will explore the key characteristics of diatoms and unveil the hidden secrets behind their beauty and multifunctionality, offering an interdisciplinary vision that celebrates the interaction between art and science. We provide concrete examples of structural materials that demonstrate how diatoms can serve as the meeting point between artistic creativity and technological innovation, inspiring new design paradigms and reshaping how we approach materials science and engineering.

## Introduction

### Introducing diatoms

Nature's beauty is rarely ornamental—it is often a direct expression of functional efficiency. Diatoms, a taxonomically diverse and ecologically foundational group of single-celled eukaryotic algae, exemplify this principle in its most refined form. Viewed under a microscope, these aquatic organisms resemble miniature opalescent jewels—encased in glass-like silica shells adorned with intricate geometric patterns and nanoscale symmetries (see Figure 1a). From space, their vast blooms trace iridescent, swirling patterns across the ocean's surface, evoking the dynamic brushstrokes of Van Gogh (see Figure 1b). Yet beneath this captivating visual façade lies one of the most vital biological engines of the planet. Diatoms are among the most prolific primary producers on Earth. Particularly abundant in nutrient-rich, high-latitude oceans, they convert dissolved carbon dioxide into organic matter while releasing oxygen—accounting for an estimated 20–30% of the global oxygen supply, a contribution rivaling that of all terrestrial rainforests [1,2]. Their ecological role as the base of marine food webs makes them critical not only to ocean health but also to the stability of the planet's climate system [3–5]. This global impact stems from a combination of extraordinary metabolic productivity and structural ingenuity. Diatoms reproduce rapidly, both sexually and asexually, allowing populations to expand explosively [6–8]. Beyond their biological function, they also leave a geological and economic legacy. Their silica shells accumulate on the ocean floor after death, forming sedimentary deposits that give rise to diatomaceous earth [9,10]. This resource has found broad application in industry, from filtration and insulation to agriculture and explosives [7,11–16].

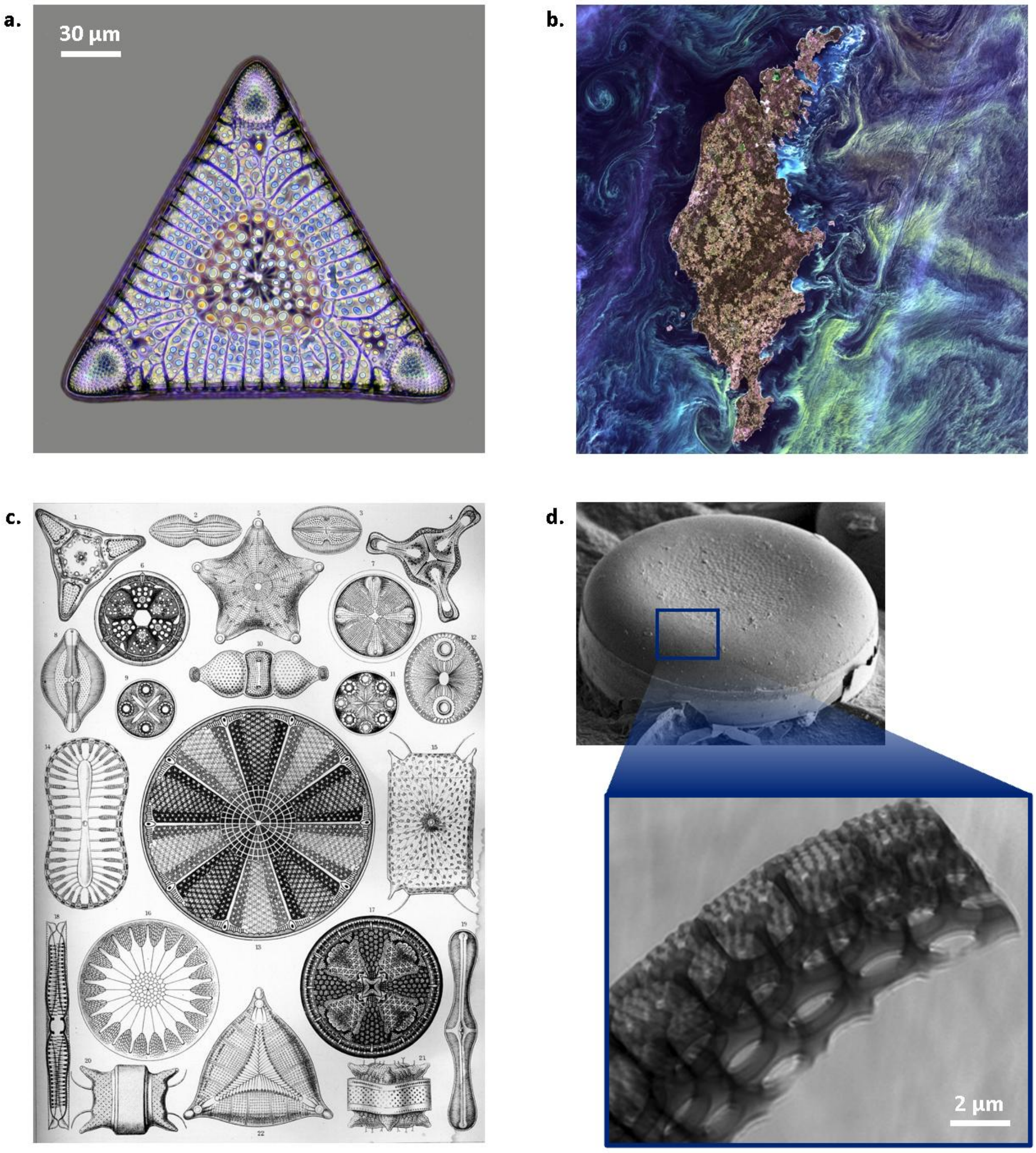

*Figure 1 – (**a.**) Optical micrograph of a diatom whose glass-like silica shell resembles opal jewelry with symmetrical decorations (source: Wikimedia Commons, Triceratium morlandii var. morlandii; reproduced under CC BY-SA 4.0.) (**b.**) Satellite image of a phytoplankton bloom, including diatoms, forming swirling surface patterns (source: Wikimedia Commons, Van Gogh from Space; reproduced under CC BY 2.0). (**c.**) Nineteenth-century illustration of diatom frustule diversity (source: Wikimedia Commons, Haeckel Diatomea 4; Public Domain). (**d.**) Scanning electron micrograph (top) of a centric diatom highlighting its Petri dish–like valve structure, and transmission electron micrograph (bottom) detailing its hierarchical architecture composed of pores, hexagonal chamber-like reinforcements, and multilayered silica walls (adapted from* [17] *under CC BY 4.0).*

## Diatoms through the lens

The visual and scientific fascination with diatoms began well before their biological sophistication was understood. In the 19th century, light microscopy provided the first window into their striking geometries [18–21]. Victorian microscopists, collectors, and naturalists were captivated by their radial and bilateral symmetries, treating diatom slides as both scientific specimens and artistic compositions (see Figure 1c). However, the limits of optical resolution meant that much of their structural intricacy remained inaccessible. It wasn't until the advent of scanning electron microscopy (SEM) in the mid-20th century that the true complexity of diatom exoskeletons could be appreciated in full detail. SEM allowed researchers to observe their surface textures, pore patterns, and multilayered morphologies with nanometer-level precision [22]. These shells are neither flat nor uniformly porous; rather, they are spatially organized into hierarchical layers, featuring nested chambers, submicron pore arrays, and ridge systems that vary across regions of the shell (see Figure 1d). Later, advanced techniques such as focused ion beam SEM (FIB-SEM), transmission electron microscopy (TEM), and atomic force microscopy (AFM) added further resolution and dimensional insight, revealing the internal architecture and even the dynamic biosilicification processes by which these structures are assembled within the living cell [17,23].

## Diatom frustule multifunctionality

As this hidden order came to light, so did its complex structure-properties relationship. The silica walls of diatoms, named frustules, revealed themselves as sophisticated multifunctional systems in which form and function are tightly interwoven [24]. Structurally, the frustule resembles a microscopic Petri dish, consisting of two overlapping halves: the larger epitheca and the smaller hypotheca. Each valve is encircled by a series of lateral girdle bands, which are added sequentially during post-mitotic growth [25] (see Figure 1d). These bands not only provide controlled flexibility and mechanical continuity but also play essential roles in cell expansion, maintaining species-specific morphology, and facilitating volumetric adjustment during development. On a macroscopic scale, the symmetry of the shell ensures a balanced distribution of mechanical loads, while its gently curved, vaulted form enhances its resistance to compressive forces [26,27]. This structural logic recalls the principles behind man-made constructions, where transparency, lightness, and strength are achieved through the strategic merge of materials and structure (see **Figure 2**a).

Though seemingly uniform at first glance, the frustule is in fact intricately ornamented and locally specialized. Thickened ribs, pore fields, spines, and other geometrical elements are hierarchically organized to optimize mechanical performance (see Figure 1d). Specifically, structural feature sizes, densities, and arrangements are not random but strategically tuned across different regions. This heterogeneous architecture provides local toughening mechanisms and energy dissipation pathways [28–30]—principles that mirror advanced mechanical design strategies into engineered materials [31–33]. Moreover, recent dynamic mechanical analyses have revealed that frustules do not behave as rigid shells but instead respond flexibly to mechanical stimuli, exhibiting compliance that evolves with repeated loading cycles—an extraordinary trait for a biologically produced mineral structure [27].

a. b.

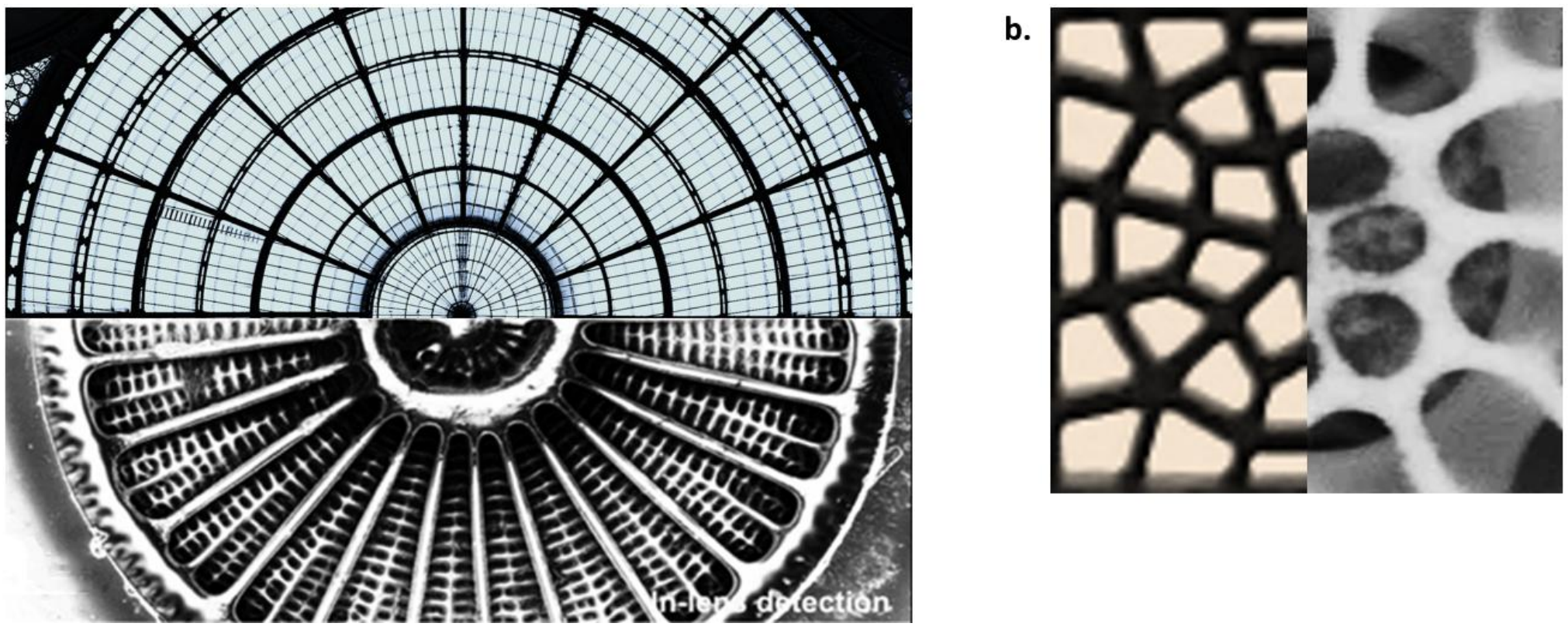

***Figure 2** – Architectural analogies between human-made and diatom structures. (**a.**) Top: glass dome of the Galleria Vittorio Emanuele II, Milan (source: Wikimedia Commons, Cupola Galleria VEII; reproduced under CC BY-SA 4.0); bottom: scanning electron micrograph of a centric diatom valve showing radially arranged ribs and pore fields (adapted from* [34]*, under CC BY 4.0). (**b.**) Left: polygonal tessellation pattern from the façade of La Zisa, Palermo—an example of Islamic architectural design related to mashrabiya latticework (source: Wikimedia Commons, La Zisa; reproduced under CC BY-SA 4.0). Right: nanoscale silica network in the frustule of the centric diatom Coscinodiscus wailesii (adapted from* [35]*, under CC BY 4.0).*

Yet the frustule's functional repertoire extends beyond mechanical defense. Its internal nanostructures interact with light in complex ways, enabling the redirection, scattering, and selectively filtering radiation across wavelengths. Such optical modulation not only enhances photosynthetic efficiency under diffuse or low-light conditions but also reflects a broader logic of environmental responsiveness [36,37]. At the same time, the spatial modulation of porosity supports selective permeability, allowing efficient exchange of gases and nutrients while maintaining structural integrity [38–40]. This ability to simultaneously transmit light, modulate spectra, and regulate flow while preserving enclosure strikingly echoes architectural strategies developed across human history to temper harsh climate [41,42] (see **Figure 2**b).

**Diatom nanotechnology**

Decoding the building logic of diatoms has exponentially increased their appeal in nanotechnology. Indeed, the biomineralization process underlying frustule morphogenesis offers a scalable, low-cost, low-impact, and highly reproducible alternative to synthetic fabrication methods for generating nanostructured materials with controlled architectures [7,12,43–46]. Unlike top-down lithographic approaches or bottom-up chemical synthesis, which typically require high energy input, hazardous solvents, or expensive instrumentation [47,48], diatoms construct complex, hierarchical silica structures under ambient conditions using only light, carbon dioxide, water, and trace nutrients. This biological process occurs intracellularly in membrane-bound silica deposition vesicles, where polymerization of silicic acid is guided by a genetically encoded molecular toolkit—including silaffins, long-chain polyamines, and silacidins—that ensures precise spatial and structural control over the resulting biosilica [49]. Additionally, because frustule formation is tightly coupled to the cell cycle, and diatoms can reproduce rapidly through both mitotic and sexual pathways, large populations can be grown in controlled culture environments with minimal infrastructure. Photobioreactors, for instance, can support dense algal growth while enabling continuous or batch harvesting of biosilica shells with high material uniformity (i.e. algal productivity can reach 70–150 million tons per hectare per year [50]), a clear advantage over natural diatomaceous earth.

Optical technologies have been among the first to capitalize on these features. Thanks to their periodic pore arrays and multiscale symmetry, diatom valves act as natural photonic crystals, capable of manipulating light in ways comparable to engineered nanostructures but at a fraction of the cost and fabrication complexity [51]. Individual frustules have been shown to focus, diffract, and guide light, producing wavelength-dependent hotspots that suggest immediate use as microlenses or passive filters in micro-optoelectronic systems [36,52,53]. Because their production does not require lithographic processing or cleanroom conditions, these biosilica elements can be integrated into optical platforms with minimal post-processing, offering a scalable alternative to traditional components in sensors, optical filters, and photonic chips.

The potential of silica, however, becomes most evident when its properties are deliberately expanded through functionalization [43,54] (see **Figure 3**a). Several approaches have been developed to nanoengineer the structure and chemistry of diatom frustules beyond their native state, acting either post-synthetically or during biosynthesis [45,55]. Surface functionalization techniques, for instance, can be applied after harvesting to confer selective chemical reactivity or affinity for specific targets. Alternatively, metabolic insertion of tailored precursors during growth enables in situ doping of the silica matrix, introducing catalytic or optoelectronic properties directly within the forming frustule. Beyond these chemical modifications, genetic engineering offers another promising route [56]. By modifying genes that encode biosilicification proteins it becomes possible to modulate frustule morphology directly within the living cell. This approach not only ensures uniformity and scalability but introduces the possibility of inheritable traits, enabling continuous production of functionally enhanced biosilica through standard cultivation.

These strategies have opened new avenues across multiple technological domains. In the biomedical field, modified diatom frustules have been explored as multifunctional drug delivery platforms, capable not only of gradual release but also of targeted localization, imaging contrast, or triggered activation [13,50,57] (see **Figure 3**b). Their tunable porosity and surface chemistry enable precise control over molecule loading and release profiles, while the rigid silica framework offers mechanical stability and protection for the payload. At the same time, their intrinsic biocompatibility facilitates safe integration into biological environments, supporting applications in both therapy and diagnostics. In the energy sector, hybrid diatom-derived structures have shown promise as photoactive templates for solar-driven catalysis, including hydrogen generation and dye-sensitized solar cells, where their hierarchical porosity and light-scattering behavior enhance photon capture and interfacial reactions [58].

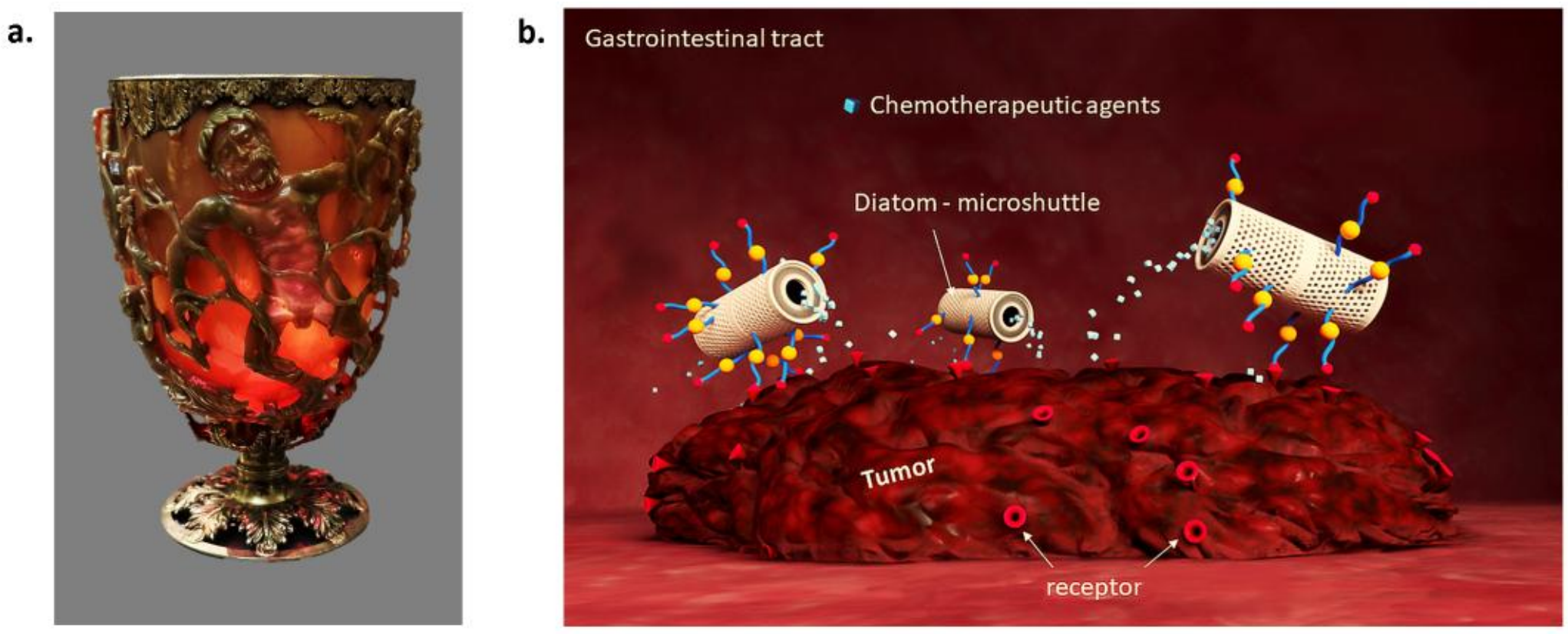

***Figure 3*** *- Evolution of silica nanotechnologies across time. (**a.**) Late Roman dichroic glass vessel (Lycurgus Cup, 4th century CE), where embedded metallic nanoparticles confer wavelength-dependent optical effects (source: Wikimedia Commons, Front of 4th century CE Roman Lycurgus Cup, British Museum (1958,1202.1); adapted under CC BY-SA 4.0). (**b.**) Contemporary nanotechnology: functionalized diatom frustules developed as multifunctional drug delivery systems, where surface engineering and photoactivation enable targeted therapeutic action (source: adapted from* [59]*, under CC BY 4.0).*

### Diatoms as models for Nature-driven innovation at all scales

In recent years, research on diatoms has moved beyond their role as biological foundries of silica nanostructures toward a deeper appreciation of the design intelligence embedded into their architecture. Once regarded mainly as a convenient source of ready-made nanoscale features, the frustule is now recognized as a paradigmatic example of multifunctional design to mimic [60]: a structure that integrates mechanical resilience, optical modulation, selective permeability, and ecological efficiency within a single hierarchical framework [24]. Such integration of diverse functions is rarely matched in synthetic systems [61,62], underscoring the value of diatoms as blueprints for rethinking material and structural design from a bioinspired, multiscale perspective, where geometry—rather than chemistry alone—emerges as a key driver of innovation [63–65]. This shift fits within a broader interdisciplinary trend in which progress increasingly relies on translating natural principles into advanced engineering solutions [66,67].

The opportunity to translate natural design strategies into engineering practice has been unlocked by rapid advances in science and technology [68–70]. High-resolution microscopy and spectroscopic methods now enable the systematic mapping of biological architectures across scales—from nanoscale pore networks to macroscale geometries—and allow the reconstruction of detailed 3D models [17,71], while computational tools—ranging from molecular simulations to continuum modeling—have made it possible to unravel how distinct functions such as mechanical resilience, light manipulation, or fluid transport emerge from hierarchical organization [28,29,40,72–77]. In parallel, additive manufacturing and self-assembly techniques provide the means to turn these insights into tangible structures, whose complexity increasingly echoes the sophistication of biological systems [68,78–81]. The result is a creative landscape where vision and feasibility evolve together: Nature teaches us what is possible, while technology demonstrates how it can be mimicked [82].

Building on this momentum, the following section provides a concise overview of architected materials and biomimetics, providing the conceptual framework needed to situate diatoms within the current “materials-by-design” paradigm [83]. The discussion then narrows to diatom-inspired structural materials, highlighting the most significant technological solutions developed to date. Together, these examples reveal how biological insight, engineering tools, and creative vision are converging to generate new discoveries. The chapter ends by outlining current challenges and future perspectives for diatom-inspired architected material design, pointing to the opportunities that arise at the intersection of biology, engineering, design, and art.

## Architected materials and biomimetics: a conceptual framework

### Embedding architecture in materials

Architecture has always been central to the way humans achieve efficiency and functionality. A telling case is the evolution of the wheel: while early versions were carved as heavy, solid disks of stone or wood, subsequent designs redistributed matter into rims, hubs, and lightweight supports, replacing bulk with void. Structural performance in this case derived not only from the intrinsic properties of the material, but from its spatial organization—strength and stiffness increased even as mass decreased (interestingly, diatoms now provide inspiration for innovations even in this domain [84,85]). The same principle extends from component-scale structures to engineering materials: when solid matter and air—or contrasting solid phases—are deliberately arranged in space according to a predefined design logic, entirely new property combinations become accessible. This is the foundation of what are broadly termed architected materials, systems whose mechanical response is determined primarily by the multiscale architecture, rather than by composition alone [70,86,87].

Since the term "architected" encompasses the merger of material and structure, the observation scale becomes crucial. Architected materials are conventionally defined at the mesoscale: their characteristic features are larger than those of conventional microstructures (such as grains in metals) but smaller than the overall component (at least in one characteristic dimension), typically spanning from tens of micrometers to millimeters [70,86,87] (see **Figure 4**). In this sense, they bridge the traditional divide between material microstructure and component structure, filling the gap that historically separated materials science from mechanical engineering. Today, this definition is expanding, as advances in nanotechnologies extend the concept of architecture down to the nanoscale [87].

The interest in architected materials stems from the evolving demands of modern engineering, where efficiency, multifunctionality, safety, smart capabilities, and sustainability are increasingly prioritized [88]. Conventional approaches, limited to tuning composition or microstructure, have proven insufficient to meet these requirements, as many applications reveal the intrinsic trade-offs of traditional materials. Enhancing one property often comes at the expense of another— as can be seen, for instance, in the classic compromise between strength and toughness in load-bearing structures [89], or between stiffness and electrical conductivity in conductors [90]. To meet such challenges, architecture at intermediate length scales has been recognized as a new degree of freedom in materials design, enabling unprecedented property combinations and expanding regions in material-property space that were previously inaccessible [90,91]. Therefore, research on architected materials remains highly active today: they promise lightweight efficiency, damage-tolerance, and multifunctionality that directly respond to market pull, regulatory requirements, and societal pressures for sustainable innovation.

At present, the most widespread industrial examples are laminated composites, cellular foams, and honeycombs, valued for their simplicity of fabrication even though they remain far from optimal for advanced applications [68–70]. Yet progress in this domain is hindered by the intrinsic complexity of their structure–process–property–performance relationships [92]. Managing of this complexity requires a multiscale engineering perspective, as different physical phenomena dominate at distinct characteristic length scales, and optimized designs can only emerge from systemic strategies that integrate mechanics, physics, chemistry, biology, and computation across multiple orders of magnitude [67].

To navigate this complexity while avoiding lengthy trial-and-error design cycles, biomimetics provides a powerful complementary route: by distilling principles from natural architectures already refined through evolution, it offers ready-made strategies to accelerate the discovery and development of next-generation architected materials [93]. Therefore, its main features are described below.

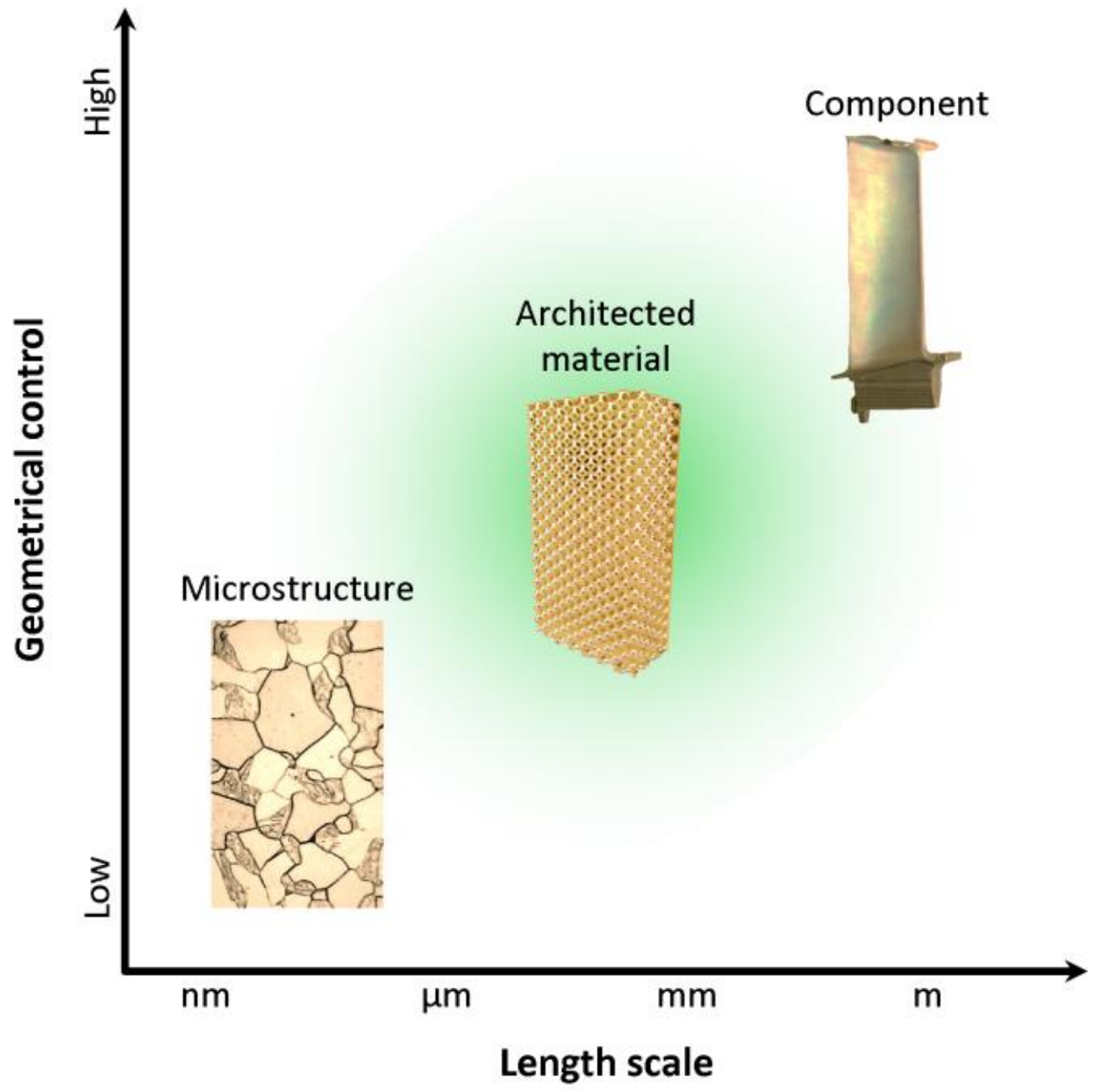

***Figure 4*** *- Architected materials bridge the scale between microstructural features and macroscopic components, enabling geometrical control across multiple length scales (images adapted from Wikimedia Commons).*

## Biomimetic design

The principles underlying architected materials resonate strongly with strategies long refined in Nature. As exemplified by diatom frustules, natural materials achieve levels of efficiency and multifunctionality that are rarely attainable in engineering practice. Proteins, polysaccharides, and minerals serve as the fundamental biological building blocks from which higher-order design elements emerge [94]. At the primary level, these include fibrils and platelets, which are subsequently assembled into fibrous bundles, helicoidal architectures, tubular channels, layered stacks, cellular lattices, overlapping tessellations, and sutured joints [65]. These motifs are integrated through specialized interfaces that regulate adhesion, mediate load transfer, and deflect cracks, while functional gradients and spatial heterogeneities finely adjust local performance to environmental and mechanical demands [64,95]. When combined into hierarchical frameworks, they generate material systems whose properties transcend the simple sum of their constituents [96]. Additionally, this multiscale logic enables biological structures to integrate functions that in engineered materials are typically antagonistic [28]. Natural systems can therefore be regarded as archetypes of architected materials, and biomimetics emerges from this awareness as the discipline that abstracts their organizational principles and translates them into technological innovation [82].

Rather than reproducing biological materials in their original composition, form, or dimensions, biomimetics aims to emulate selected functional principles through engineering processes and design methodologies, with the objective of generating solutions optimized for specific technological challenges. In this way it offers a pathway to bypass the prolonged trial-and-error cycles that typically arise from the intricate structure–property relationships of hierarchical, multiscale materials. Indeed, the translation of natural strategies into engineering practice inevitably entails a shift in chemical constituents, processing methods, and characteristic length scales. Whereas natural materials are built from biopolymers and minerals under mild conditions of temperature and pressure, engineered counterparts are more commonly fabricated from metals, ceramics, and synthetic polymers through energy-intensive processes that operate with levels of precision fundamentally different from those achieved by biology [97]. This divergence in material bases, manufacturing environments and length scales alters the physics governing performance, such that many emergent phenomena observed in natural systems cannot be directly reproduced in artificial ones [98]. Biomimetics therefore acts as a selective filter, identifying which natural strategies can be abstracted and reformulated into engineering materials, while discarding or reinterpreting those that cannot be meaningfully transferred across base constituents, processes, and scales.

This becomes particularly evident in the case of diatom frustules. Their submicron porosity and multilayered silica walls confer a unique combination of low weight, high strength, and local compliance at the microscale, but when similar geometries are transposed into polymeric or metallic lattices at millimetric scales, the mechanisms of energy dissipation and crack arrest change fundamentally, governed no longer by nanoscale flaw tolerance [99] but by conventional bulk fracture mechanics . Likewise, the periodic pore arrays that act as diffraction gratings and photonic lenses in the visible spectrum lose their optical function when scaled up for lightweight architectural components [98]. Yet their structural role in distributing loads and increasing stiffness-to-weight efficiency remains intact, while the multilayered porous structure can instead be exploited for functions such as enhanced permeability, selective filtration, or mass transport. Even in fluid dynamics, the riblets and surface ornamentations that regulate nutrient exchange and flow at the low Reynolds numbers characteristic of microalgal environments [40,98] cannot be directly transferred to macroscopic systems, where inertial phenomena dominate; nevertheless, the same geometric logic can inspire engineered surfaces for drag reduction, flow redirection, or temporary fluid storage in industrial and biomedical devices.

To overcome such intrinsic limits, biomimetics is increasingly converging with data-driven methodologies. Advances in Artificial Intelligence now allow the vast design space of natural architectures to be systematically analyzed and abstracted into predictive frameworks. Deep learning models can identify recurring structural motifs, explore parametric variations beyond those found in Nature, and anticipate performance trade-offs with unprecedented efficiency [100–104]. When coupled with multiscale simulations, these tools accelerate the shift from descriptive analogy to optimization-driven design pipelines, expanding the reach of biomimetics from selective imitation to a data-augmented, predictive paradigm for architected materials [100–104].

## Emerging diatom-inspired structural materials

Diatom frustules rank among the most strong and tough lightweight biological architectures known in Nature [29,105] (see **Figure 5**), and their diversity reflects extraordinary adaptation capabilities to distinct environmental pressures. Research on diatom-inspired structural materials therefore focuses on two fundamental questions. The first is how the relationship between diatoms' structure and properties can be decoded to identify optimal design strategies that maximize properties-to-weight ratios beyond what conventional material approaches achieve. This involves isolating the role of hierarchy or specific geometric motifs—such as cellular layers or porosity—in governing mechanical response under various loading conditions. The second is how diatom-inspired architectures can be modified and tuned to meet targeted functional requirements, for instance by introducing gradients, controlled heterogeneities, or hierarchical variations that adjust stiffness or energy absorption.

**Decoding structure–property relationships for optimal mechanical design**

To answer the first question, early studies showed how diatom-inspired architectures can improve silica mechanical behavior through hierarchical design at the smallest scales. Luo and Greer [105] fabricated synthetic valves from an artificial silica precursor (i.e. POSS: cyclohexyl polyhedral oligomeric silsesquioxanes) with geometries inspired by *Coscinodiscus* frustules using two-photon lithography direct laser writing, and demonstrated, through in-situ three-point bending and fracture tests on FIB-milled micro-beams inside an SEM, that the mimicked hierarchical architecture reproduces the ability to slow down crack propagation and enhance toughness even when the chemistry differs from natural biosilica (see **Figure 5**b). Their work established that geometry can be isolated as a decisive factor in resisting fracture in glass-like materials, offering a benchmark for evaluating structure–property relations independently of composition and highlighting how structural motifs could inform the design of lightweight, damage-resistant ceramics. In parallel, molecular dynamics and mesoscale simulations showed that the introduction of hierarchical levels into otherwise brittle foils markedly increases strain capacity as well as toughness and strength [28,106,107] (see **Figure 5**c). This improvement arises from cooperative shearing and crack-arresting mechanisms, which replace the catastrophic fracture typically observed in single-level structures**.** These results emphasize the functional role of hierarchy in silica architectures and underline its potential for developing mechanically robust, yet lightweight, architected materials.

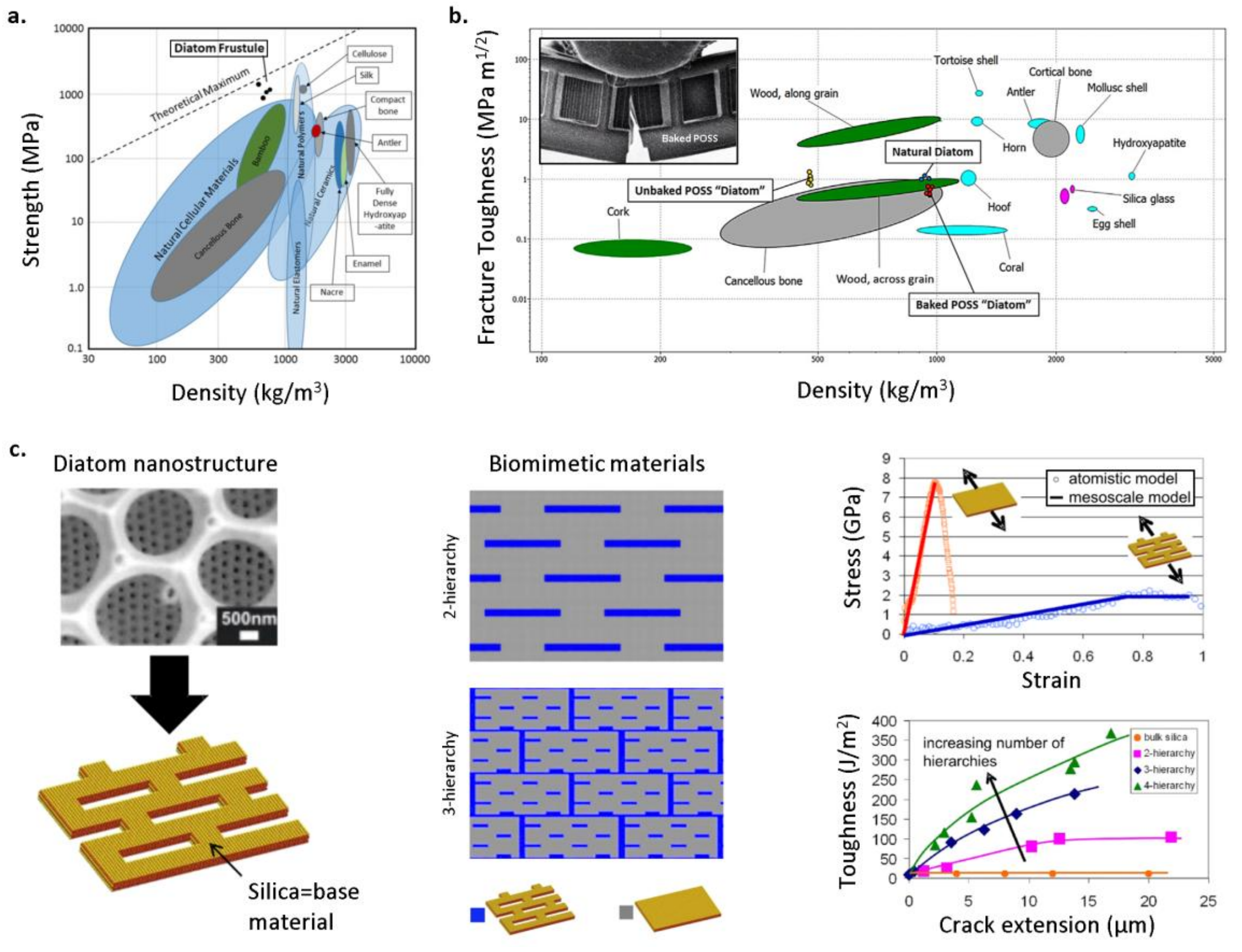

***Figure 5** – Mapping and decoding structure–property relationships of diatoms for biomimetic design. (**a.**) Ashby plot of natural materials comparing strength versus density, highlighting the outstanding performance of diatom frustules (adapted from* [29]*). (**b.**) Ashby plot of fracture toughness versus density showing natural diatom biosilica alongside bio-mimicked frustule architectures fabricated by two-photon lithography (adapted with permission from* [105]*). (**c.**) Computational design of a hierarchical architected material inspired by diatom frustules, where successive structural levels dramatically increase defect-tolerance and crack-resistance compared to monolithic silica (adapted from* [28] *under CC BY-NC-ND 3.0).*

Subsequent research moved toward more general design abstractions using polymers. Specifically, Gutiérrez et al. [108] modeled *Coscinodiscus* valves and their unit cells with finite element simulations and validated their manufacturability through 3D direct laser writing of polymer replicas at the microscale, showing that basal and middle layers of the sandwich-like diatom architecture dominate global stiffness, while pore size and shape tune local stress distributions and can trigger stiffness losses if excessively elongated. This work positioned diatom-inspired geometries as transferable templates for microdevice engineering, paving the way for applications ranging from stiffness-optimized materials to biomedical micro-implants where structure–property relations can be exploited for tailored mechanical performance.

Alongside static performance, the structure-properties relationship has also been investigated from the dynamics point of view, since diatoms in Nature experience not only static compression but also jackhammer-like vibrational loads during copepod predation [109]. Abdusatorov et al. [110] analyzed frustules of *Coscinodiscus* and *Synedra acus* by means of finite element simulations to establish how morphological parameters—including external dimensions, wall thickness, pore size, density, and elastic modulus—influence natural frequencies and mode shapes. Their results provided correlations that link geometry to eigenfrequency response, offering a quantitative basis for transferring frustule-inspired design into vibration-resistant microstructures. Gutierrez et al. [111] extended this approach by combining simulation and experimental validation to demonstrate how variations in pore arrangement and layer distribution alter deformation and vibrational modes. These findings confirmed that dynamic properties, just like static ones, can be systematically tuned by exploiting architectural motifs observed in diatom shells. More recently, Andresen et al. [72] investigated dome-like structures inspired by diatom architectures, showing through numerical models that specific stiffening patterns such as ribs or comb-like elements significantly shift eigenfrequencies without adding mass. When compared with conventional thickness optimization, these bioinspired patterns provided superior vibration control at constant weight. Together, these studies underline that frustule morphology not only enhances strength and stiffness but also governs dynamic stability, thereby offering new strategies to design lightweight components capable of avoiding resonance while maintaining high efficiency in properties-to-weight ratios.

**Tuning material architectures for targeted functional requirements**

In addressing the second question, research has explored how diatom-inspired architectures can be tuned through gradients, heterogeneities, and parametric modifications to meet targeted functional requirements. A first line of investigation has focused on stiffness optimization. Breish et al. [112] introduced two bio-inspired optimization workflows that abstract the uneven and graded material distribution typical of diatom frustules. The first workflow, inspired by *Auliscus intermedius*, generates continuous plate structures with optimized boundaries and local thickness variations, mimicking the species' surface thickening strategy. The second workflow, inspired by *Triceratium sp.*, produces three-dimensional cellular solids with graded distributions of cell size and wall thickness. Using implicit modeling and topology optimization, they demonstrated that such graded designs outperform conventional solid–void lattices by redistributing material only where mechanically most effective. This approach highlighted how evolutionary solutions for material economy in diatoms can be translated into engineering strategies that maximize stiffness-to-weight ratios at scales suitable for structural applications. In parallel, Linnemann et al. [113] developed a stress-adaptive stiffening strategy that abstracts diatom-like ribs, combs, and local thickenings into parametric surface stiffeners on plates and shells (see Figure 6a). Using purely computational models at the component scale (structural steel assigned uniformly, constant mass), they showed that introducing rib patterns aligned with stress fields can reduce static deflection by up to 93% compared to reference plates of equal weight. Importantly, the automated workflow generalizes to curved geometries and different boundary conditions, making it broadly applicable. Complementing this static approach, Andresen and Basri [114] focused on dynamic stiffness tuning by deforming 3D structures along their mode shapes. Through extensive numerical studies, they demonstrated that this strategy can increase eigenfrequencies by more than 200% without adding mass, thereby improving vibration resistance at constant weight. These studies show that stiffness can be systematically tuned by structural features rather than material substitution, providing scalable, material-independent design principles that extend diatom-inspired concepts beyond static strength to include adaptive stiffening and vibration control, thereby reinforcing the role of diatom architectures as a blueprint for next-generation lightweight architected materials.

A second line of investigation has focused on mechanical energy absorption tuning. Musenich et al. [115] proposed a data-driven biomimetic approach inspired by the graded wall thickness observed in the honeycomb-like structure of *Coscinodiscus* frustules (see Figure 6b). The study focused on the out-of-plane compression response of elastomeric honeycombs at the millimeter scale, fabricated from thermoplastic polyurethane (TPU) via additive manufacturing. A dataset of thousands of bioinspired geometries was generated by varying wall-thickness distributions, and finite element analyses were validated against quasi-static compression tests on 3D-printed samples. Machine learning algorithms were then trained to link geometric features to mechanical properties, particularly absorbed energy per unit peak force. The optimized structures exhibited up to a 150% increase in performance compared to conventional constant-thickness honeycombs. This work demonstrated how functional gradients typical of diatom shells can be exploited by predictive tools to design tunable energy-absorbing materials, providing scalable guidelines for protective devices and lightweight impact-resistant components. More recently, Musenich et al. [116] extended the approach to a practical application by developing a diatom-inspired helmet liner concept. Starting from the multilayer architecture of *Coscinodiscus* frustules, they designed periodic biomimetic multilayered structures at the millimeter scale, again using TPU as the base material and fused filament fabrication for prototyping. Finite element analysis, analytical modeling, and quasistatic compression tests on 3D-printed samples were combined to assess mechanical behavior and optimize geometry. Results showed that adding diatom-inspired porous layers significantly improved energy absorption compared to simple honeycombs, and parametric optimization further increased absorbed elastic energy about 20%. Although the study focused on mechanical properties, it also highlighted the potential for extending diatom-inspired materials to multifunctional helmets where energy absorption, fluid-dynamic performance, and breathability are integrated into the liner. This demonstrates how diatom-inspired architected materials can move beyond proof-of-concept geometries toward application-ready designs, offering a pathway to next-generation protective equipment with enhanced safety and multifunctionality.

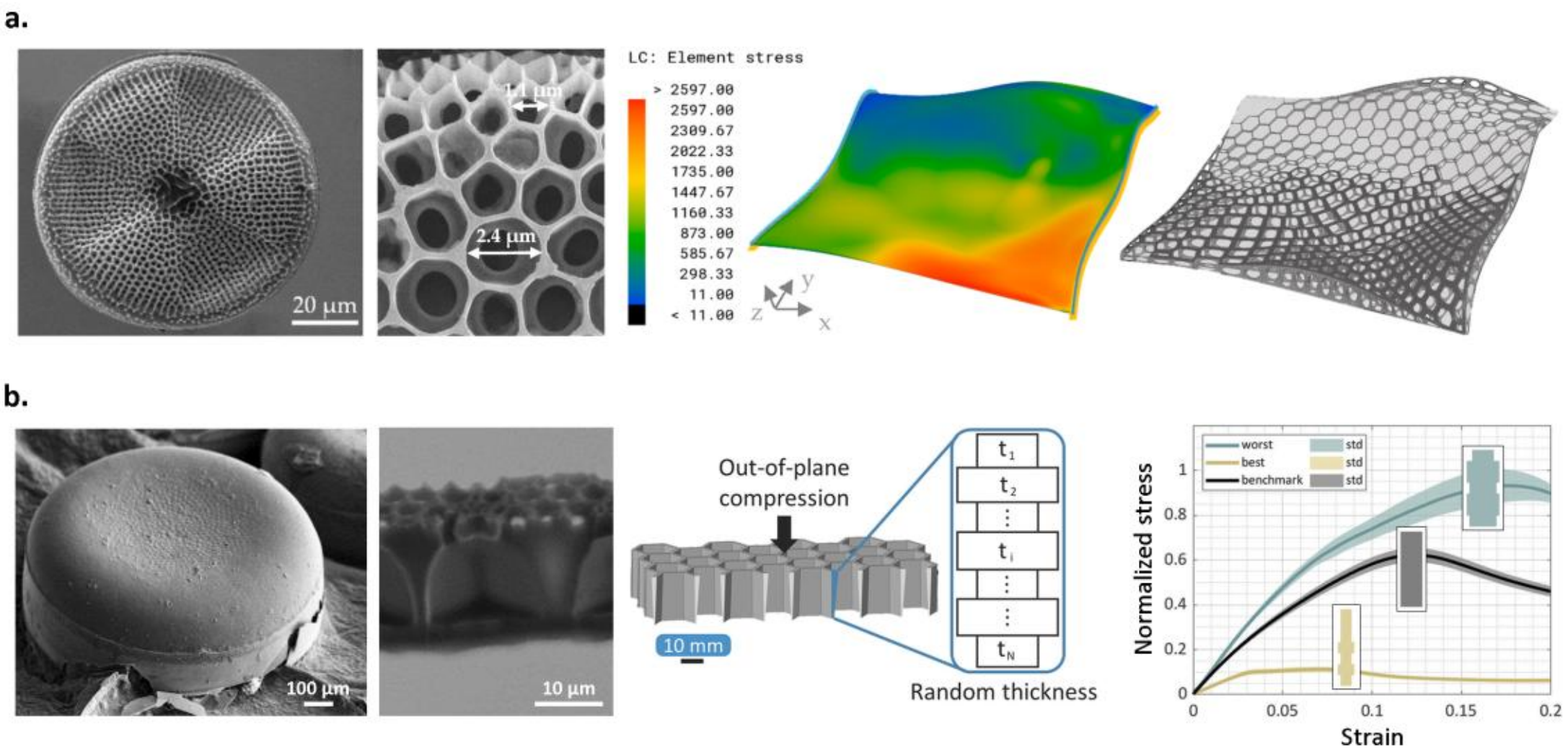

***Figure 6*** *– Diatom-inspired strategies for bending stiffness and energy absorption tuning. (**a.**) SEM images highlight the graded pore organization of diatoms and an example of a bio-inspired stress-adaptive stiffening (adapted from* [113] *under CC BY 4.0). (**b.**) Biomimetic honeycombs with tunable mechanical properties: starting from the thickness gradients found in natural diatom, 3D-printed structures reveal how geometrical tailoring controls energy absorption profiles (diatom images adapted from* [17] *under CC BY 4.0, biomimetic honeycomb and related results adapted with permission from* [115]*).*

## Towards multifunctional biomimetic materials

Since diatom frustules are inherently multifunctional, biomimetic research is shifting from focusing solely on mechanical optimization toward exploring multifunctional design strategies. However, this is still an emerging field, with only a few studies explicitly addressing how diatom-inspired architectures can integrate different physical roles within the same structural material system.

A first example is provided by Musenich et al. [117], who addressed the challenge of linking mechanical performance with fluid-dynamic behavior in a unified biomimetic framework (see Figure 7a). Taking inspiration from the multilayered shell architecture of *Coscinodiscus*, they developed mesoscale prototypes fabricated from polymer resins by high-resolution additive manufacturing. The biomimetic strategy abstracted layer-specific structural features of the frustule, which were then systematically varied in 3D-printed models. Numerical analyses, analytical formulations, and mechanical tests revealed that the basal plate, characterized by thick ribs and dense pores, was primarily responsible for flexural efficiency, while the honeycomb-like layer contributed significantly to global stiffness and buckling resistance. In contrast, the upper porous layer played a dominant role in modulating fluid flow, enabling more homogeneous distribution across the structure. The results showed that the basic *Coscinodiscus* architecture is inherently optimized to simultaneously maximize structural and fluid-dynamic efficiency, and that this multifunctional balance is preserved even when transposed into different materials and scales. As a consequence, the study points to the potential of diatom-inspired designs for multifunctional architected materials in applications such as biomedical scaffolds and fluidic devices, where the integration of mechanical stability with controlled transport is essential.

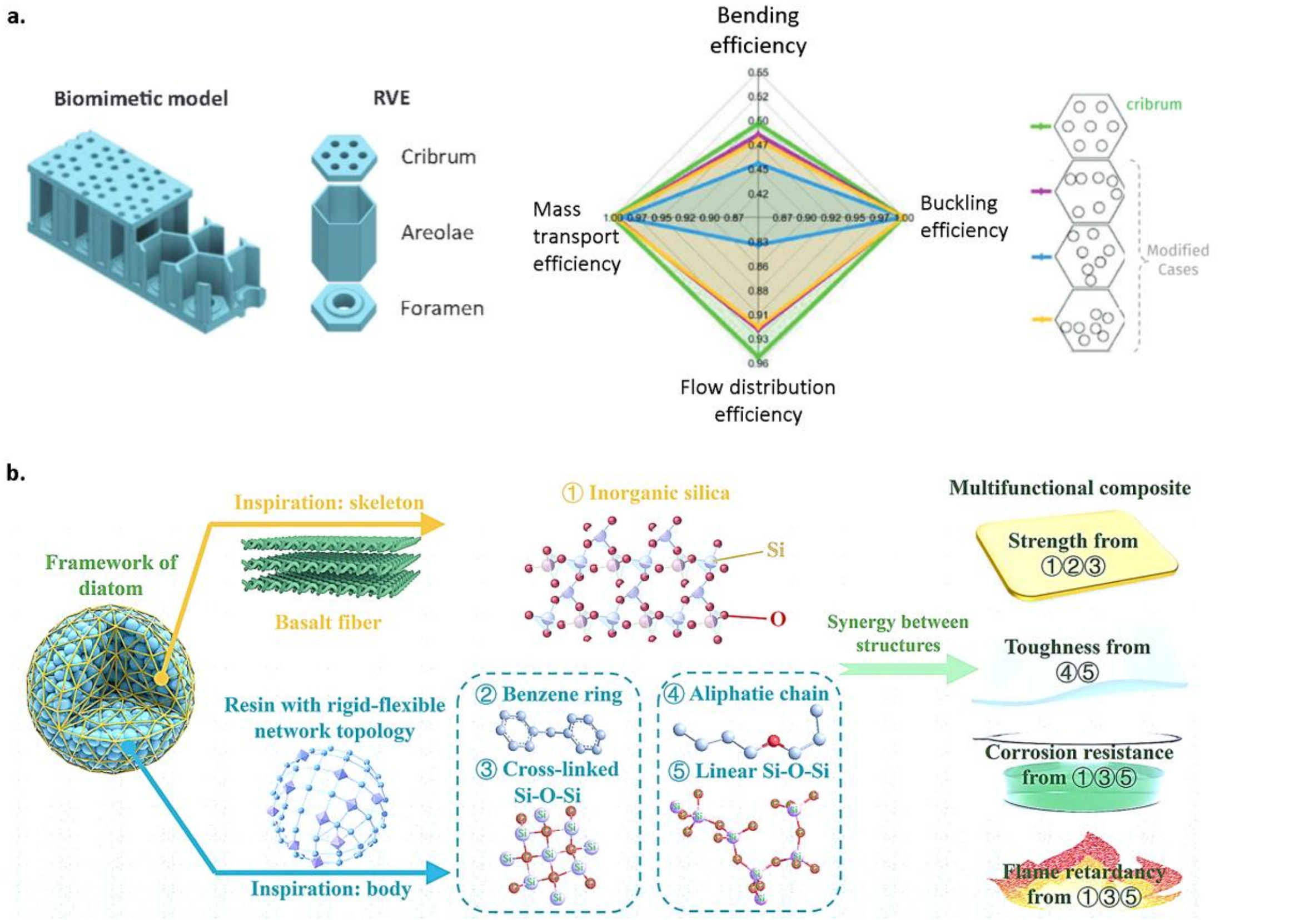

***Figure 7*** *- Diatom-inspired multifunctional design strategies. (**a.**) Biomimetic modeling of Coscinodiscus frustules: the hierarchical substructures are abstracted into representative volume elements and evaluated in terms of bending, buckling, mass transport, and flow distribution efficiencies. Radar plots comparing natural and modified cases highlight the evolutionary optimization of the frustule architecture for multifunctionality. (Adapted from* [117] *under CC BY 4.0). (**b.**) Bioinspired composite mimicking the dual inorganic–organic organization of diatoms: The synergy between materials and structure overcomes the strength–toughness trade-off, while also conferring corrosion resistance and flame retardancy. (Reproduced with permission from* [118]*).*

A second example is offered by Xu et al. [118], who addressed the challenge of simultaneously achieving strength, toughness, corrosion resistance, and flame retardancy in fiber-reinforced polymer composites—properties that are often mutually exclusive (see Figure 7b). Inspired by the composite nature of diatom frustules, which combine inorganic silica with organic components in a rigid–flexible network, they designed a basalt fiber-reinforced epoxy composite using a bioinspired network topology approach. Basalt fibers, largely composed of silica, provided the continuous inorganic framework, while an epoxy resin containing both carbon- and silicon-based chains was synthesized to mimic the organic–inorganic hybrid logic of frustules. Mechanical testing showed significant improvements compared to both conventional epoxy composites and a commercial multifunctional reference system, with tensile, flexural, and impact properties all enhanced while maintaining resistance to crack propagation. Beyond mechanical performance, the diatom-inspired design also enabled excellent resistance to acid, alkali, and salt corrosion, while flame tests demonstrated self-extinguishing behavior and reduced flammability, confirming effective flame-retardant performance. This study illustrated how diatom-inspired principles can be abstracted into structural composites that reconcile traditionally conflicting requirements, paving the way for safer and more durable structures.

## Conclusions and future perspectives

Diatoms exemplify how Nature achieves multifunctionality by merging strength, toughness, permeability, light modulation, and adaptability within a single hierarchical structure. Building on this awareness, biomimetic research in materials' mechanics has already yielded important insights for innovation: hierarchical geometries can be abstracted into scalable design principles for enhancing toughness, optimizing stiffness, resisting vibrations, tuning energy absorption. More recently, first attempts have shown that these principles can also be extended to multifunctional domains, where structural performance is coupled with fluid transport, chemical resistance, or flame retardancy. Nevertheless, most of these advances remain at the proof-of-concept stage, limited to specific materials, scales, or isolated functions. Diatom-inspired design approaches are rapidly expanding beyond structural mechanics, with similar design logics emerging in disciplines such as optics and photonics [76,119–121], energy conversion [122,123], water remediation and harvesting [124,125]. Therefore, the key challenge now is to progress from individual demonstrations toward a systematic design framework where diatom-inspired strategies can be generalized, optimized, and implemented in multi-objective engineering applications.

A promising path forward lies in identifying the design grammar encoded in frustules—the set of rules refined by evolution to balance efficiency with adaptability. By formalizing how pores, ribs, and layers are organized to control mechanical robustness, fluid exchange, or optical response, researchers can transform this biological intelligence into generative design frameworks. Such frameworks would enable inverse design, where desired property profiles are computationally translated into optimized architectures. Advances in materials informatics and generative artificial intelligence make this vision increasingly realistic: biological structures can be digitized into parameterized datasets, machine learning models can capture correlations between geometry and function, and generative algorithms can propose new morphologies that extend beyond those observed in Nature. In this way, diatoms can evolve from biological curiosities into design toolkits for creating sustainable, multifunctional, and adaptive architected materials across domains ranging from lightweight structures to photonics and biomedical devices [126].

Within this natural grammar, symmetry occupies an attractive potential for innovation. In *Coscinodiscus* and related genera, cyclic and dihedral symmetry groups govern the arrangement of ribs and pore lattices, ensuring isotropic load distribution and uniform light modulation [127,128]. These strategies resonate strongly with modern architected materials, where symmetric motifs are exploited in lattices, photonic crystals, and materials to program wave propagation, energy dispersion, and structural efficiency. Beyond engineering, the power of symmetry echoes through culture and art:

much like the diatoms, for instance, M.C. Escher manipulated repetition, periodicity, and transformation to create works of striking balance and beauty. In both cases, whether in microscopic shells shaped by evolution or in human creations shaped by imagination, symmetry emerges as a universal principle through which complexity is ordered, functionality is unlocked, and elegance is achieved.

## References

1. Field, C.B., Behrenfeld, M.J., Randerson, J.T., and Falkowski, P. (1998) Primary Production of the Biosphere: Integrating Terrestrial and Oceanic Components. *Science (80-. ).*, **281** (5374), 237–240.
2. Nelson, D.M., Tréguer, P., Brzezinski, M.A., Leynaert, A., and Quéguiner, B. (1995) Production and dissolution of biogenic silica in the ocean: Revised global estimates, comparison with regional data and relationship to biogenic sedimentation. *Global Biogeochem. Cycles*, **9** (3), 359–372.
3. Benoiston, A.-S., Ibarbalz, F.M., Bittner, L., Guidi, L., Jahn, O., Dutkiewicz, S., and Bowler, C. (2017) The evolution of diatoms and their biogeochemical functions. *Philos. Trans. R. Soc. B Biol. Sci.*, **372** (1728), 20160397.
4. Falkowski, P.G., Katz, M.E., Knoll, A.H., Quigg, A., Raven, J.A., Schofield, O., and Taylor, F.J.R. (2004) The Evolution of Modern Eukaryotic Phytoplankton. *Science (80-. ).*, **305** (5682), 354–360.
5. Armbrust, E.V. (2009) The life of diatoms in the world's oceans. *Nature*, **459** (7244), 185–192.
6. Poulíčková, A., Mann, D.G., and Mann, D.G. (2019) Diatom Sexual Reproduction and Life Cycles, in *Diatoms: Fundamentals and Applications*, Wiley, pp. 245–272.
7. Kröger, N., and Poulsen, N. (2008) Diatoms—From Cell Wall Biogenesis to Nanotechnology. *Annu. Rev. Genet.*, **42** (1), 83–107.
8. Malviya, S., Scalco, E., Audic, S., Vincent, F., Veluchamy, A., Poulain, J., Wincker, P., Iudicone, D., de Vargas, C., Bittner, L., Zingone, A., and Bowler, C. (2016) Insights into global diatom distribution and diversity in the world's ocean. *Proc. Natl. Acad. Sci.*, **113** (11), E1516–E1525.
9. Losic, D., and Korunic, Z. CHAPTER 10. Diatomaceous Earth, A Natural Insecticide for Stored Grain Protection: Recent Progress and Perspectives, pp. 219–247.
10. Zahajská, P., Opfergelt, S., Fritz, S.C., Stadmark, J., and Conley, D.J. (2020) What is diatomite? *Quat. Res.*, **96** (2004), 48–52.
11. Vinayak, V., Joshi, K.B., and Sarma, P.M. (2019) Diafuel™ (Diatom Biofuel) vs Electric Vehicles, a Basic Comparison: A High Potential Renewable Energy Source to Make India Energy Independent, in *Diatoms: Fundamentals and Applications*, Wiley, pp. 537–582.
12. Gordon, R., Losic, D., Tiffany, M.A., Nagy, S.S., and Sterrenburg, F.A.S. (2009) The Glass Menagerie: diatoms for novel applications in nanotechnology. *Trends Biotechnol.*, **27** (2), 116–127.
13. Maher, S., Kumeria, T., Aw, M.S., and Losic, D. (2018) Diatom Silica for Biomedical Applications: Recent Progress and Advances. *Adv. Healthc. Mater.*, **7** (19).
14. Sharma, N., Simon, D.P., Diaz-Garza, A.M., Fantino, E., Messaabi, A., Meddeb-Mouelhi, F., Germain, H., and Desgagné-Penix, I. (2021) Diatoms Biotechnology: Various Industrial Applications for a Greener Tomorrow. *Front. Mar. Sci.*, **8** (February).
15. Boons, R., Siqueira, G., Grieder, F., Kim, S., Giovanoli, D., Zimmermann, T., Nyström, G., Coulter, F.B., and Studart, A.R. (2023) 3D Bioprinting of Diatom-Laden Living Materials for Water Quality Assessment. *Small*, **19** (50), 1–13.
16. Wang, Z., Gong, D., and Cai, J. (2021) Diatom Frustule Array for Flow-Through Enhancement of Fluorescent Signal in a Microfluidic Chip. *Micromachines*, **12** (9), 1017.
17. Zgłobicka, I., Gluch, J., Liao, Z., Werner, S., Guttmann, P., Li, Q., Bazarnik, P., Plocinski, T.,

Witkowski, A., and Kurzydlowski, K.J. (2021) Insight into diatom frustule structures using various imaging techniques. *Sci. Rep.*, **11** (1), 14555.

18. Kociolek, J.P., Williams, D.M., Hamsher, S., Miller, S., and Li, J. (2024) Studies on type material from Kützing's diatom collection VIII. Species assigned to the genera Epithemia Brébisson ex Kützing and Rhopalodia O. Müller. *Diatom Res.*, **39** (4), 159–186.
19. Kahn, R. (1995) C. G. Ehrenberg's Concept of the Diatoms. *Arch. für Protistenkd.*, **146** (2), 109–116.
20. Dolan, J.R. (2024) The protists of Ernst Haeckel's Kunstformen der Natur. *Protist*, **175** (6), 126069.
21. Smith, H.L. (1886) A Contribution to the Life History of the Diatomaceæ. *Proc. Am. Soc. Microsc.*, **8**, 30–66.
22. Tiffany, M.A., and Nagy, S.S. (2019) The Beauty of Diatoms, in *Diatoms: Fundamentals and Applications*, Wiley, pp. 33–42.
23. Manoylov, K., and Ghobara, M. (2021) Introduction for a Tutorial on Diatom Morphology, in *Diatom Morphogenesis*, Wiley, pp. 1–18.
24. De Tommasi, E., Gielis, J., and Rogato, A. (2017) Diatom Frustule Morphogenesis and Function: a Multidisciplinary Survey. *Mar. Genomics*, **35** (2017), 1–18.
25. Round, F.E., Crawford, R.M., and Mann, D.G. (1990) *The Diatoms. Biology and Morphology of the Genera.*, Cambridge University Press.
26. Hamm, C.E., Merkel, R., Springer, O., Jurkojc, P., Maier, C., Prechtel, K., and Smetacek, V. (2003) Architecture and material properties of diatom shells provide effective mechanical protection. *Nature*, **421** (6925), 841–843.
27. Cvjetinovic, J., Luchkin, S.Y., Statnik, E.S., Davidovich, N.A., Somov, P.A., Salimon, A.I., Korsunsky, A.M., and Gorin, D.A. (2023) Revealing the static and dynamic nanomechanical properties of diatom frustules—Nature's glass lace. *Sci. Rep.*, **13** (1), 5518.
28. Sen, D., and Buehler, M.J. (2011) Structural hierarchies define toughness and defect-tolerance despite simple and mechanically inferior brittle building blocks. *Sci. Rep.*, **1** (1), 35.
29. Aitken, Z.H., Luo, S., Reynolds, S.N., Thaulow, C., and Greer, J.R. (2016) Microstructure provides insights into evolutionary design and resilience of Coscinodiscus sp. frustule. *Proc. Natl. Acad. Sci.*, **113** (8), 2017–2022.
30. Huang, W., Restrepo, D., Jung, J., Su, F.Y., Liu, Z., Ritchie, R.O., McKittrick, J., Zavattieri, P., and Kisailus, D. (2019) Multiscale Toughening Mechanisms in Biological Materials and Bioinspired Designs. *Adv. Mater.*, **31** (43), 1–37.
31. Yang, X.-Y., Chen, L.-H., Li, Y., Rooke, J.C., Sanchez, C., and Su, B.-L. (2017) Hierarchically porous materials: synthesis strategies and structure design. *Chem. Soc. Rev.*, **46** (2), 481–558.
32. Gibson, Lorna J., Ashby, M.F. (1988) *Cellular Solids: Structure and Propertes.*, Cambridge University Press.
33. Barthelat, F. (2015) Architectured materials in engineering and biology: fabrication, structure, mechanics and performance. *Int. Mater. Rev.*, **60** (8), 413–430.
34. Ferrara, M.A., Dardano, P., De Stefano, L., Rea, I., Coppola, G., Rendina, I., Congestri, R., Antonucci, A., De Stefano, M., and De Tommasi, E. (2014) Optical Properties of Diatom Nanostructured Biosilica in Arachnoidiscus sp: Micro-Optics from Mother Nature. *PLoS One*, **9** (7), e103750.
35. De Tommasi, E., Congestri, R., Dardano, P., De Luca, A.C., Managò, S., Rea, I., and De Stefano, M. (2018) UV-shielding and wavelength conversion by centric diatom nanopatterned frustules. *Sci. Rep.*, **8** (1), 16285.
36. Ghobara, M.M., Ghobara, M.M., Mazumder, N., Vinayak, V., Reissig, L., Gebeshuber, I.C., Tiffany, M.A., Gordon, R., and Gordon, R. (2019) On Light and Diatoms: A Photonics and Photobiology Review, in *Diatoms: Fundamentals and Applications*, Wiley, pp. 129–189.
37. Mcheik, A., Cassaignon, S., Livage, J., Gibaud, A., Berthier, S., and Lopez, P.J. (2018) Optical Properties of Nanostructured Silica Structures From Marine Organisms. *Front. Mar. Sci.*, **5** (APR).

38. Rosengarten, G., and Herringer, J.W. CHAPTER 2. Interactions of Diatoms with Their Fluid Environment, pp. 14–54.
39. Arrieta, J., Jeanneret, R., Roig, P., and Tuval, I. (2020) On the fate of sinking diatoms: the transport of active buoyancy-regulating cells in the ocean. *Philos. Trans. R. Soc. A Math. Phys. Eng. Sci.*, **378** (2179), 20190529.
40. Guasto, J.S., Rusconi, R., and Stocker, R. (2012) Fluid Mechanics of Planktonic Microorganisms. *Annu. Rev. Fluid Mech.*, **44** (1), 373–400.
41. Taki, A., and Kumari, H. (2023) Examining Mashrabiya's Impact on Energy Efficiency and Cultural Aspects in Saudi Arabia. *Sustainability*, **15** (13), 10131.
42. Fraternali, F., Babilio, E., Nazifi Charandabi, R., Germano, G., Luciano, R., and Spagnuolo, G. (2024) Dynamic origami solar eyes with tensegrity architecture for energy harvesting Mashrabiyas. *Appl. Eng. Sci.*, **19** (August), 100190.
43. Halas, N.J. (2008) Nanoscience under Glass: The Versatile Chemistry of Silica Nanostructures. *ACS Nano*, **2** (2), 179–183.
44. Losic, D., Mitchell, J.G., and Voelcker, N.H. (2009) Diatomaceous Lessons in Nanotechnology and Advanced Materials. *Adv. Mater.*, **21** (29), 2947–2958.
45. Ragni, R., Cicco, S.R., Vona, D., and Farinola, G.M. (2018) Multiple Routes to Smart Nanostructured Materials from Diatom Microalgae: A Chemical Perspective. *Adv. Mater.*, **30** (19), 1–23.
46. Liu, X., Zhang, F., Jing, X., Pan, M., Liu, P., Li, W., Zhu, B., Li, J., Chen, H., Wang, L., Lin, J., Liu, Y., Zhao, D., Yan, H., and Fan, C. (2018) Complex silica composite nanomaterials templated with DNA origami. *Nature*, **559** (7715), 593–598.
47. Liddle, J.A., and Gallatin, G.M. (2016) Nanomanufacturing: A Perspective. *ACS Nano*, **10** (3), 2995–3014.
48. van Assenbergh, P., Meinders, E., Geraedts, J., and Dodou, D. (2018) Nanostructure and Microstructure Fabrication: From Desired Properties to Suitable Processes. *Small*, **14** (20).
49. Hildebrand, M., Lerch, S.J.L., and Shrestha, R.P. (2018) Understanding Diatom Cell Wall Silicification—Moving Forward. *Front. Mar. Sci.*, **5** (APR), 1–19.
50. Rea, I., Terracciano, M., and De Stefano, L. (2017) Synthetic vs Natural: Diatoms Bioderived Porous Materials for the Next Generation of Healthcare Nanodevices. *Adv. Healthc. Mater.*, **6** (3).
51. Fuhrmann, T., Landwehr, S., El Rharbi-Kucki, M., and Sumper, M. (2004) Diatoms as living photonic crystals. *Appl. Phys. B*, **78** (3–4), 257–260.
52. De Tommasi, E. (2016) Light Manipulation by Single Cells: The Case of Diatoms. *J. Spectrosc.*, **2016**, 1–13.
53. Ragni, R., Cicco, S., Vona, D., Leone, G., and Farinola, G.M. (2017) Biosilica from diatoms microalgae: smart materials from bio-medicine to photonics. *J. Mater. Res.*, **32** (2), 279–291.
54. Bayda, S., Adeel, M., Tuccinardi, T., Cordani, M., and Rizzolio, F. (2019) The History of Nanoscience and Nanotechnology: From Chemical–Physical Applications to Nanomedicine. *Molecules*, **25** (1), 112.
55. Sun, X., Zhang, M., Liu, J., Hui, G., Chen, X., and Feng, C. (2024) The Art of Exploring Diatom Biosilica Biomaterials: From Biofabrication Perspective. *Adv. Sci.*, **11** (6), 1–21.
56. Trofimov, A.A., Pawlicki, A.A., Borodinov, N., Mandal, S., Mathews, T.J., Hildebrand, M., Ziatdinov, M.A., Hausladen, K.A., Urbanowicz, P.K., Steed, C.A., Ievlev, A. V., Belianinov, A., Michener, J.K., Vasudevan, R., and Ovchinnikova, O.S. (2019) Deep data analytics for genetic engineering of diatoms linking genotype to phenotype via machine learning. *npj Comput. Mater.*, **5** (1), 67.
57. Delalat, B., Sheppard, V.C., Rasi Ghaemi, S., Rao, S., Prestidge, C.A., McPhee, G., Rogers, M.-L., Donoghue, J.F., Pillay, V., Johns, T.G., Kröger, N., and Voelcker, N.H. (2015) Targeted drug delivery using genetically engineered diatom biosilica. *Nat. Commun.*, **6** (1), 8791.
58. Sun, X.W., Zhang, Y.X., and Losic, D. (2017) Diatom silica, an emerging biomaterial for energy conversion and storage. *J. Mater. Chem. A*, **5** (19), 8847–8859.

59. Delasoie, J., Schiel, P., Vojnovic, S., Nikodinovic-Runic, J., and Zobi, F. (2020) Photoactivatable Surface-Functionalized Diatom Microalgae for Colorectal Cancer Targeted Delivery and Enhanced Cytotoxicity of Anticancer Complexes. *Pharmaceutics*, **12** (5), 480.
60. Kooistra, W.H.C.F., and Pohl, G. (2015) Diatom Frustule Morphology and its Biomimetic Applications in Architecture and Industrial Design, pp. 75–102.
61. Ferreira, A.D.B.L., Nóvoa, P.R.O., and Marques, A.T. (2016) Multifunctional Material Systems: A state-of-the-art review. *Compos. Struct.*, **151**, 3–35.
62. Christodoulou, L., and Venables, J.D. (2003) Multifunctional material systems: The first generation. *JOM*, **55** (12), 39–45.
63. Jia, Z., Deng, Z., and Li, L. (2022) Biomineralized Materials as Model Systems for Structural Composites: 3D Architecture. *Adv. Mater.*, **34** (20).
64. Liu, Z., Meyers, M.A., Zhang, Z., and Ritchie, R.O. (2017) Functional gradients and heterogeneities in biological materials: Design principles, functions, and bioinspired applications. *Prog. Mater. Sci.*, **88**, 467–498.
65. Naleway, S.E., Porter, M.M., McKittrick, J., and Meyers, M.A. (2015) Structural Design Elements in Biological Materials: Application to Bioinspiration. *Adv. Mater.*, **27** (37), 5455–5476.
66. Oxman, N. (2016) Age of Entanglement. *J. Des. Sci.*
67. Cranford, S.W., and Buehler, M.J. (2012) *Biomateriomics*, Springer Netherlands, Dordrecht.
68. Surjadi, J.U., and Portela, C.M. (2025) Enabling three-dimensional architected materials across length scales and timescales. *Nat. Mater.*, **24** (4), 493–505.
69. Xia, X., Spadaccini, C.M., and Greer, J.R. (2022) Responsive materials architected in space and time. *Nat. Rev. Mater.*, **7** (9), 683–701.
70. Schaedler, T.A., and Carter, W.B. (2016) Architected Cellular Materials. *Annu. Rev. Mater. Res.*, **46** (1), 187–210.
71. BALL, A.D., JOB, P.A., and WALKER, A.E.L. (2017) SEM-microphotogrammetry, a new take on an old method for generating high-resolution 3D models from SEM images. *J. Microsc.*, **267** (2), 214–226.
72. Andresen, S., Linnemann, S.K., Ahmad Basri, A.B., Savysko, O., and Hamm, C. (2024) Natural Frequencies of Diatom Shells: Alteration of Eigenfrequencies Using Structural Patterns Inspired by Diatoms. *Biomimetics*, **9** (2), 85.
73. Nguyen, H., and Fauci, L. (2014) Hydrodynamics of diatom chains and semiflexible fibres. *J. R. Soc. Interface*, **11** (96), 20140314.
74. Herringer, J.W., Lester, D., Dorrington, G.E., and Rosengarten, G. (2019) Can diatom girdle band pores act as a hydrodynamic viral defense mechanism? *J. Biol. Phys.*, **45** (2), 213–234.
75. De Tommasi, E., Rea, I., Mocella, V., Moretti, L., De Stefano, M., Rendina, I., and De Stefano, L. (2010) Multi-wavelength study of light transmitted through a single marine centric diatom. *Opt. Express*, **18** (12), 12203.
76. Shapturenka, P., Isaac Zakaria, N., Birkholz, F., and Gordon, M.J. (2023) Extending the diatom's color palette: non-iridescent, disorder-mediated coloration in marine diatom-inspired nanomembranes. *Opt. Express*, **31** (13), 21658.
77. De Stefano, L., Rea, I., Rendina, I., De Stefano, M., and Moretti, L. (2007) Lensless light focusing with the centric marine diatom Coscinodiscus walesii. *Opt. Express*, **15** (26), 18082.
78. Zglobicka, I., Chmielewska, A., Topal, E., Kutukova, K., Gluch, J., Krüger, P., Kilroy, C., Swieszkowski, W., Kurzydlowski, K.J., and Zschech, E. (2019) 3D Diatom–Designed and Selective Laser Melting (SLM) Manufactured Metallic Structures. *Sci. Rep.*, **9** (1), 19777.
79. Colombo, P., and Franchin, G. (2021) Printing glass in the nano. *Nat. Mater.*, **20** (11), 1454–1456.
80. Zhu, Z., Gao, D., Huang, Z., Chang, W., Wu, B., Zhang, K., Sun, M., Song, H., Ritchie, R.O., Wang, T., Huang, W., and Zhou, H. (2025) Cryogenic 3D printing of damage tolerant hierarchical porous ceramics. *Int. J. Extrem. Manuf.*, **7** (4), 045002.
81. Mirkhalaf, M., and Barthelat, F. (2017) Design, 3D printing and testing of architectured materials with bistable interlocks. *Extrem. Mech. Lett.*, **11**, 1–7.

82. Vincent, J.F. V (2009) Biomimetics — a review. *Proc. Inst. Mech. Eng. Part H J. Eng. Med.*, **223** (8), 919–939.
83. Olson, G.B. (2000) Designing a New Material World. *Science (80-. ).*, **288** (5468), 993–998.
84. Hamm, C., and Möller, S. (2015) ELiSE—An Integrated, Holistic Bionic Approach to Develop Optimized Lightweight Solutions for Engineering, Architecture and Design, pp. 183–194.
85. Kaiser, N., Goossens, N., Jimenez, A., Laraudogoitia, I., Psarras, S., and Tsantzalis, S. (2024) Advanced manufacturing concept of a bio-inspired reaction wheel rotor for small- and medium-sized constellation satellites. *CEAS Sp. J.*, **16** (1), 73–86.
86. Estrin, Y., Bréchet, Y., Dunlop, J., and Fratzl, P. (eds.) (2019) *Architectured Materials in Nature and Engineering*, Springer International Publishing, Cham.
87. Barthelat, F. (2015) Architectured materials in engineering and biology: fabrication, structure, mechanics and performance. *Int. Mater. Rev.*, **60** (8), 413–430.
88. Zhang, B., Xu, P., Wang, J., Hong, Z., Wang, W., and Dai, F. (2025) Overcoming the trade-off between conductivity and strength in copper alloys through undercooling. *Nat. Commun.*, **16** (1), 4978.
89. Ritchie, R.O. (2011) The conflicts between strength and toughness. *Nat. Mater.*, **10** (11), 817–822.
90. Bréchet, Y.J.M. (2013) CHAPTER 1. Architectured Materials: An Alternative to Microstructure Control for Structural Materials Design? A Possible Playground for Bio-inspiration?, pp. 1–16.
91. Fleck, N.A., Deshpande, V.S., and Ashby, M.F. (2010) Micro-architectured materials: past, present and future. *Proc. R. Soc. A Math. Phys. Eng. Sci.*, **466** (2121), 2495–2516.
92. McDowell, D.L., and Olson, G.B. (2008) Concurrent design of hierarchical materials and structures, in *Lecture Notes in Computational Science and Engineering*, pp. 207–240.
93. Vincent, J.F. V (2009) Biomimetics — a review. *Proc. Inst. Mech. Eng. Part H J. Eng. Med.*, **223** (8), 919–939.
94. Marth, J.D. (2008) A unified vision of the building blocks of life. *Nat. Cell Biol.*, **10** (9), 1015–1015.
95. Barthelat, F., Yin, Z., and Buehler, M.J. (2016) Structure and mechanics of interfaces in biological materials. *Nat. Rev. Mater.*, **1** (4), 16007.
96. Fratzl, P. (2007) Biomimetic materials research: what can we really learn from nature's structural materials? *J. R. Soc. Interface*, **4** (15), 637–642.
97. Meyers, M.A., and Chen, P.-Y. (2014) *Biological Materials Science*, Cambridge University Press.
98. Perricone, V., Santulli, C., Rendina, F., and Langella, C. (2021) Organismal Design and Biomimetics: A Problem of Scale. *Biomimetics*, **6** (4), 56.
99. Gao, H., Ji, B., Jäger, I.L., Arzt, E., and Fratzl, P. (2003) Materials become insensitive to flaws at nanoscale: Lessons from nature. *Proc. Natl. Acad. Sci.*, **100** (10), 5597–5600.
100. Gu, G.X., Chen, C.-T., Richmond, D.J., and Buehler, M.J. (2018) Bioinspired hierarchical composite design using machine learning: simulation, additive manufacturing, and experiment. *Mater. Horizons*, **5** (5), 939–945.
101. Guo, K., Yang, Z., Yu, C.-H., and Buehler, M.J. (2021) Artificial intelligence and machine learning in design of mechanical materials. *Mater. Horizons*, **8** (4), 1153–1172.
102. Buehler, M.J. (2024) MechGPT, a Language-Based Strategy for Mechanics and Materials Modeling That Connects Knowledge Across Scales, Disciplines, and Modalities. *Appl. Mech. Rev.*, **76** (2).
103. Luu, R.K., and Buehler, M.J. (2023) Materials Informatics Tools in the Context of Bio-Inspired Material Mechanics. *J. Appl. Mech.*, **90** (9).
104. Lee, J., Park, D., Lee, M., Lee, H., Park, K., Lee, I., and Ryu, S. (2023) Machine learning-based inverse design methods considering data characteristics and design space size in materials design and manufacturing: a review. *Mater. Horizons*, **10** (12), 5436–5456.
105. Luo, S., and Greer, J.R. (2018) Bio-Mimicked Silica Architectures Capture Geometry, Microstructure, and Mechanical Properties of Marine Diatoms. *Adv. Eng. Mater.*, **20** (9), 1–9.
106. Garcia, A.P., Sen, D., and Buehler, M.J. (2011) Hierarchical Silica Nanostructures Inspired by

Diatom Algae Yield Superior Deformability, Toughness, and Strength. *Metall. Mater. Trans. A*, **42** (13), 3889–3897.
107. Garcia, A.P., Pugno, N., and Buehler, M.J. (2011) Superductile, Wavy Silica Nanostructures Inspired by Diatom Algae. *Adv. Eng. Mater.*, **13** (10), 405–414.
108. Gutiérrez, A., Guney, M.G., Fedder, G.K., and Dávila, L.P. (2018) The role of hierarchical design and morphology in the mechanical response of diatom-inspired structures via simulation. *Biomater. Sci.*, **6** (1), 146–153.
109. Michels, J., Vogt, J., and Gorb, S.N. (2012) Tools for crushing diatoms – opal teeth in copepods feature a rubber-like bearing composed of resilin. *Sci. Rep.*, **2** (1), 465.
110. Abdusatorov, B., Salimon, A.I., Bedoshvili, Y.D., Likhoshway, Y. V., and Korsunsky, A.M. (2020) FEM exploration of the potential of silica diatom frustules for vibrational MEMS applications. *Sensors Actuators A Phys.*, **315**, 112270.
111. Gutiérrez, A., Gordon, R., and Dávila, L.P. (2017) Deformation Modes And Structural Response Of Diatom Frustules. *J. Mater. Sci. Eng. with Adv. Technol.*, **15** (2), 105–134.
112. Breish, F., Hamm, C., and Kienzler, R. (2023) Diatom-inspired stiffness optimization for plates and cellular solids. *Bioinspir. Biomim.*, **18** (3), 036004.
113. Linnemann, S.K., Friedrichs, L., and Niebuhr, N.M. (2024) Stress-Adaptive Stiffening Structures Inspired by Diatoms: A Parametric Solution for Lightweight Surfaces. *Biomimetics*, **9** (1), 46.
114. Andresen, S., and Ahmad Basri, A.B. (2024) Diatom-Inspired Structural Adaptation According to Mode Shapes: A Study on 3D Structures and Software Tools. *Biomimetics*, **9** (4), 241.
115. Musenich, L., Stagni, A., Derin, L., and Libonati, F. (2024) Tunable Energy Absorption in 3D-Printed Data-Driven Diatom-Inspired Architected Materials. *ACS Mater. Lett.*, **6** (6), 2213–2222.
116. Musenich, L., Strozzi, L., Avalle, M., and Libonati, F. (2025) D-HAT: A Diatom-Inspired Structure for a Helmet Concept Against Trauma. *Adv. Intell. Syst.*, **7** (4), 1–9.
117. Musenich, L., Origo, D., Gallina, F., Buehler, M.J., and Libonati, F. (2025) Revealing Diatom-Inspired Materials Multifunctionality. *Adv. Funct. Mater.*, **35** (8), 1–12.
118. Xu, Q., Zhang, S., Cheng, S., Zhao, Y., Wang, K., Jin, T., Long, T., Jiang, M., and Liu, P. (2025) Bioinspired network topology of multifunctional basalt fiber-reinforced polymer composite for building safety: strengthening, toughening, corrosion resistance, and flame retardancy. *Polymer (Guildf).*, **333** (June), 128633.
119. Xie, X., Huang, Y., Yang, Z., Li, A., and Zhang, X. (2024) Diatom Cribellum-Inspired Hierarchical Metamaterials: Unifying Perfect Absorption Toward Subwavelength Color Printing. *Adv. Mater.*, **36** (33), 1–10.
120. Li, A., Zhao, X., Duan, G., Anderson, S., and Zhang, X. (2019) Diatom Frustule-Inspired Metamaterial Absorbers: The Effect of Hierarchical Pattern Arrays. *Adv. Funct. Mater.*, **29** (22), 1–7.
121. Xie, P., Chen, Z., Xu, J., Xie, D., Wang, X., Cui, S., Zhou, H., Zhang, D., and Fan, T. (2019) Artificial ceramic diatoms with multiscale photonic architectures via nanoimprint lithography for CO2 photoreduction. *J. Am. Ceram. Soc.*, **102** (8), 4678–4687.
122. Song, J., Fan, Y., Cheng, Z., Wang, F., Shi, X., Xu, J., Zhang, J., Yi, H., Shuai, Y., and Zhang, H. (2025) Biomimetic low carbonization efficient solar-driven thermochemical energy storage reactor design inspired by the diatoms' superior photosynthesis capacity. *Energy Convers. Manag.*, **323** (PA), 119224.
123. Liao, G., Qin, J., Ren, L., Ren, Z., Xie, J., Cui, D., Cheng, N., Han, W., Du, Y., and Qi, X. (2025) Diatom-Inspired 3D Hierarchical Liquid Metal Sponge for Flexible Photoelectrochemical Photodetectors. *ACS Appl. Mater. Interfaces*, **17** (30), 43808–43819.
124. Pan, J., Yu, X., Dong, J., Zhao, L., Liu, L., Liu, J., Zhao, X., and Liu, L. (2021) Diatom-Inspired TiO 2 -PANi-Decorated Bilayer Photothermal Foam for Solar-Driven Clean Water Generation. *ACS Appl. Mater. Interfaces*, **13** (48), 58124–58133.
125. Chen, Y., Ji, Y., Li, X., Hou, K., and Cai, Z. (2025) Diatoms Inspired Green Janus Fabric for Efficient Fog Harvesting. *Adv. Sustain. Syst.*, **9** (1), 1–10.
126. Buehler, M.J. (2023) Diatom-inspired architected materials using language-based deep

learning: Perception, transformation and manufacturing. 1–9.
127. Pappas, J.L., Tiffany, M.A., and Gordon, R. (2021) The Uncanny Symmetry of Some Diatoms and Not of Others: A Multi-Scale Morphological Characteristic and a Puzzle for Morphogenesis, in *Diatom Morphogenesis*, Wiley, pp. 19–67.
128. Ghobara, M.M., Tiffany, M.A., Gordon, R., and Reissig, L. (2021) Diatom Pore Arrays' Periodicities and Symmetries in the Euclidean Plane: Nature Between Perfection and Imperfection, in *Diatom Morphogenesis*, Wiley, pp. 117–158.